\documentclass[letterpaper,twocolumn,table,xcdraw, 10pt]{article}
\usepackage{usenix-2020-09}
\usepackage{subcaption}
\usepackage{multirow}
\usepackage[linesnumbered,lined,ruled,vlined]{algorithm2e}
\usepackage{listings}
\usepackage{amsfonts}
\usepackage[colorinlistoftodos,prependcaption]{todonotes}
\usepackage{xspace}
\usepackage[switch]{lineno}
\usepackage{enumitem}
\usepackage{xcolor}
\usepackage{comment}
\usepackage[frozencache,cachedir=.]{minted}
\usepackage{fontawesome5}
\usepackage[affil-it]{authblk}

\AtEndEnvironment{listing}{\vspace{-10pt}}

\usepackage{tikz}
\usepackage{amsmath}

\usepackage{xcolor}
\newcommand\AB[1]{\textcolor{red}{AB: #1}}

\usepackage{xurl}

\begin{document}

\newcommand{\flowa}{{$\mathbb{F}_A$}\xspace}
\newcommand{\flowb}{{$\mathbb{F}_B$}\xspace}
\newcommand{\flowexit}{{$\mathbb{F}_{EXIT}$}\xspace}
\newcommand{\BHn}{\texttt{BH[n]}\xspace}
\newcommand{\Bxevict}{\texttt{Bx\_evict}\xspace}
\newcommand{\Biprobe}{\texttt{Bx\_prime}\xspace}
\newcommand{\talt}{\texttt{t\_alt}\xspace}
\newcommand{\tprimary}{\texttt{t\_primary}\xspace}

\newcommand{\Bev}{\(\mathcal{B}_{ev}\)\xspace}
\newcommand{\Bipred}{\texttt{Bi\_pred}\xspace}
\newcommand{\tsafe}{\texttt{t\_safe}\xspace}
\newcommand{\tleak}{\texttt{t\_leak}\xspace}
\newcommand{\Bchard}{\texttt{Bc\_hard}\xspace}
\newcommand{\Bcexit}{\texttt{Bc\_exit}\xspace}
\newcommand{\Bcinit}{\texttt{Bc\_init}\xspace}
\newcommand{\Bcload}{\texttt{Bc\_load}\xspace}
\newcommand{\Bcfp}{\texttt{Bc\_fp}\xspace}

\definecolor{keywordcolor}{rgb}{0.0, 0.0, 1.0} 
\definecolor{commentcolor}{rgb}{0.0, 0.5, 0.0}
\definecolor{stringcolor}{rgb}{0.6, 0.2, 0.8}
\definecolor{greentext}{rgb}{0.0, 0.6, 0.0} 
\definecolor{redtext}{rgb}{0.8, 0.0, 0.0}   

\lstdefinestyle{armstyle}{
    basicstyle=\ttfamily\footnotesize,
    keywordstyle=\color{keywordcolor}\bfseries,
    commentstyle=\color{commentcolor}\itshape,
    stringstyle=\color{stringcolor},
    backgroundcolor=\color{gray!10},
    frame=single,
    numbers=left,
    numberstyle=\tiny\color{gray},
    captionpos=b,
    breaklines=true,
    showstringspaces=false,
    escapeinside={(*@}{@*)}, 
    morekeywords={MOV, LDR, ADD, STR, B, NOP, RET, BLR, ADR} 
}

\date{}


\title{\Large \bf Exploiting Inaccurate Branch History in Side-Channel Attacks}

\author[1,2]{Yuhui Zhu}
\author[1]{Alessandro Biondi}
\affil[1]{Scuola Superiore Sant'Anna}
\affil[2]{Scuola IMT Alti Studi Lucca}

\maketitle

\begin{abstract}
Modern out-of-order CPUs heavily rely on speculative execution for performance optimization, with branch prediction serving as a cornerstone to minimize stalls and maximize efficiency.
Whenever shared branch prediction resources lack proper isolation and sanitization methods, they may originate security vulnerabilities that expose sensitive data across different software contexts.

This paper examines the fundamental components of modern Branch Prediction Units (BPUs) and investigates how resource sharing and contention affect two widely implemented but underdocumented features: \emph{Bias-Free Branch Prediction} and \emph{Branch History Speculation}.
Our analysis demonstrates that these BPU features, while designed to enhance speculative execution efficiency through more accurate branch histories, can also introduce significant security risks. 
We show that these features can inadvertently modify the Branch History Buffer (BHB) update behavior and create new primitives that trigger malicious mis-speculations.

This discovery exposes previously unknown cross-privilege attack surfaces for Branch History Injection (BHI).
Based on these findings, we present three novel attack primitives: two Spectre attacks, namely {\bf Spectre-BSE} and {\bf Spectre-BHS}, and a cross-privilege control flow side-channel attack called {\bf BiasScope}.
Our research identifies corresponding patterns of vulnerable control flows and demonstrates exploitation on multiple processors. Finally, \textbf{Chimera} is presented: an attack demonstrator based on eBPF for a variant of Spectre-BHS that is capable of leaking kernel memory contents at 24,628 bit/s. 


\end{abstract}

\section{Introduction}

Microarchitectural vulnerabilities pose serious and evolving threats to modern out-of-order CPUs.
Research over the past few years has demonstrated how performance-oriented optimizations, designed to maximize pipeline efficiency, can be exploited to create a malicious transient execution environment capable of leaking sensitive data across different security contexts~\cite{298146,  hofmannSpeculationFaultModeling2023, jinOvertakeAchievingMeltdowntype2023,  katzmanGatesTimeImproving2023, mambrettiBypassingMemorySafety2021, milburnYouCannotAlways2023, purnalShowTimeAmplifyingArbitrary2023, ravichandranPACMANAttackingARM2022, schluterFetchBenchSystematicIdentification2023, shivakumarSpectreDeclassifiedReading2023, tanInvisibleProbeTiming2021,  zhangMWAITItBridging2023}.
Some findings have also revealed how micro-architectural side effects can influence the behavior of shared hardware resources, enabling data disclosure attacks in concurrent execution environments~\cite{qiuPMUSpillPerformanceMonitor2022, wuRenderingContentionChannel2022, zhangBunnyHopExploitingInstruction2023, renSeeDeadUops2021, meulemeesterSpectrEMExploitingElectromagnetic2023, tatarTLBDREnhancingTLBbased2022, rohanReverseEngineeringStream2020, gastSQUIPExploitingScheduler2023, vicarteAuguryUsingData2022, yuSynchronizationStorageChannels2023}.

Despite extensive efforts to address these issues at the hardware-software interface, our research reveals that certain underdocumented micro-architectural features, albeit intended to manage resource sharing and contention, can create new attack surfaces when interacting with other micro-architectural behaviors.
\paragraph{Exploitations.}
This paper examines \emph{history-based branch prediction} and assesses potentially vulnerable behaviors inadvertently implemented in processors.
Through extensive empirical analyses of multiple processors, we reveal how micro-architectural handling of resource contention creates distinct vulnerabilities across different processor designs.

Building on our findings, we present a series of novel attack flows that can extract secret data from separate software contexts.
First, we present \textbf{BiasScope}, a coarse-grained control flow side-channel that exploits the BPU's bias-free behavior to leak branch outcomes.
Second, we propose two Spectre-variant attacks, \textbf{Branch Status Eviction (Spectre-BSE)}, and \textbf{Branch Hisotory Speculation (Spectre-BHS)}.
These attacks build upon the concept of \emph{Branch History Injection} (BHI)~\cite{barberisBranchHistoryInjection2022}, but achieve malicious manipulation of the  Branch History Buffer (BHB) \emph{indirectly} by controlling the branch history updating mechanisms we investigated.
Since they avoid explicit branch history injection through adversary-controlled branches, these attacks naturally circumvent existing BHI mitigations on ARM processors.


While the effectiveness of our attacks heavily depends on the structure of victim code, we demonstrate vulnerable code patterns and analyze their relationship to hardware implementation characteristics.
Finally, we present \textbf{Chimera}, a demonstrator based on eBPF, representative of an \emph{end-to-end attack} with Spectre-BHS, capable of leaking kernel memory at 24,628 bit/s. In the light of recent work~\cite{wiebingInSpectreGadgetInspecting2024} that unveiled a large residual attack surface for known Spectre attacks in the Linux kernel, the availability of gadgets that can enable native Spectre-BSE and Spectre-BHS attacks should certainly not be underestimated.
Research on \emph{speculative trojans}~\cite{zhangExploringBranchPredictors2020} further exacerbates the risks associated with these attacks.


\paragraph{Contribution.}

In summary, this paper makes the following contribution:
\begin{itemize}
    \item We systematically analyze resource sharing and contention in modern branch prediction units, identifying mechanisms that can lead to exploitable behaviors.
    
    

    \item We reveal and evaluate undocumented features in modern BPUs, introducing new techniques for implicit BHB manipulation. We exploit these primitives to present novel side-channel attacks: Spectre-BSE, Spectre-BHS, and BiasScope, enabling both speculative execution attacks and control flow monitoring across privilege boundaries with all Spectre mitigations enabled.

    \item Through the development of exploitable program patterns and the Chimera end-to-end attack demonstrator based on eBPF, we highlight the importance of systematic analysis in uncovering potential security vulnerabilities in hardware and software designs.


\end{itemize}

\section{Background and Related Work} \label{background}


\subsection{Branch Prediction} \label{sec:branch-pred}
To minimize branch resolution latency and determine the next fetch address before branch resolution, processors employ a \textbf{Branch Prediction Unit (BPU)} that makes educated guesses based on the historical behavior of branches.
The BPU primarily predicts two critical properties of a branch: the \emph{target address} (indirect branches) and the \emph{taken/not-taken direction} (conditional branches).
To enable predictions, the BPU implements dedicated caches to store learned branch behaviors.
The \textbf{Branch Target Buffer (BTB)}~\cite{laljaReducingBranchPenalty1988,changImprovingBranchPrediction1997,changTargetPredictionIndirect1997,driesenCascadedPredictorEconomical1998} caches target addresses, enabling early instruction fetch redirection even before branch decode completion.
The \textbf{Pattern History Table (PHT)}~\cite{yehTwolevelAdaptiveTraining1991,yehAlternativeImplementationsTwolevel1992,kiseBimodeBranchPredictor2005} aids in predicting conditional branches by tracking their past outcomes using \emph{saturation counters}.

\paragraph{History-Based Branch Prediction.}
While early branch predictors relied on simple Program Counter-based indexing to correlate branch addresses with their targets, research has shown this approach to be insufficient since branch outcomes often depend on the control flow context established by preceding branches.
Contemporary BPUs leverage this insight by implementing history-based prediction policies that capture correlations among branches.

The majority of modern BPUs introduce \textbf{Branch History Buffer (BHB)}~\cite{yehTwolevelAdaptiveTraining1991,yehAlternativeImplementationsTwolevel1992} to maintain a record of recent branch outcomes on the execution path, typically as a shift register of taken/not-taken bits, which length is pre-defined.
Some implementations employ an enhanced variant of BHB known as the \textbf{Path History Register (PHR)}~\cite{nairDynamicPathbasedBranch1995,seznecAnalysisOGEometricHistory2005}.
Unlike the canonical BHB, which only records taken/not-taken outcomes, the PHR maintains a complete jumping path by storing multi-bit footprints that encode both the source and target addresses of each \emph{taken} branch, thus providing more distinctive signatures for different control flow paths.
In PHR, the historical information is combined with the branch address to create sophisticated indexing functions for the BTB and PHT~\cite{changImprovingBranchPrediction1997}.
This approach maintains separate prediction entries for each unique control flow context, significantly improving prediction accuracy by capturing path-specific branch behaviors.
Recent studies~\cite{armSpectreBHBSpeculativeTarget2023,yavarzadehHalfHalfDemystifyingIntels2023,yavarzadehPathfinderHighResolutionControlFlow2024,barberisBranchHistoryInjection2022,299742} have provided detailed insights into the BHB updating mechanisms in modern processors.

\paragraph{Tags in BPU implementations.}

\emph{Tags} are unique identifiers in cache memory that verify the presence of requested information.
Differently from index functions, which in set-associative caches may map multiple aliasing addresses to the same cache set, tags \emph{enable unique identification} of cached elements within each set. 
Upon a cache query, tags help filter out aliased entries that share the same index and allows the cache to signal a \emph{query miss} when no matching tag is found.

Since branch prediction only guides speculative execution without affecting the architectural state, tag fields in BTB and PHT can be optional.
While many early works~\cite{laljaReducingBranchPenalty1988,changImprovingBranchPrediction1997,driesenCascadedPredictorEconomical1998,yehTwolevelAdaptiveTraining1991,yehAlternativeImplementationsTwolevel1992,kiseBimodeBranchPredictor2005} omitted tags from their designs based on this rationale, some researchers~\cite{kaeliBranchHistoryTable1991,sprangleAgreePredictorMechanism1997} introduced tags in their full-associative BTB/PHT implementations to prevent aliasing-induced mispredictions.

\begin{figure}
\centering
\includegraphics[width=\linewidth]{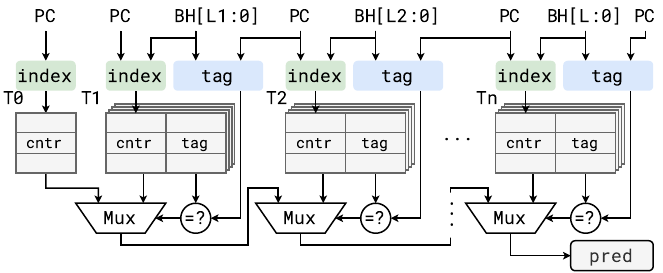}
\caption{TAGE branch predictor.}
\label{fig:tage}
\end{figure}

Tags play a central role in \emph{TAgged GEometric history length} (TAGE), the state-of-the-art branch predictor design~\cite{seznec201164,seznecNewCaseTAGE2011,seznec64KbytesITTAGEIndirect2011}.
As illustrated in Fig.~\ref{fig:tage}, TAGE consists of multiple prediction tables: several tagged tables that use different combinations of the Program Counter (PC) and branch histories with geometrically increasing length (\texttt{BH[L1:0]}, \texttt{BH[L2:0]},...) for indexing, and an untagged \emph{base predictor} (\texttt{T0}) that relies solely on PC-based indexing.
During prediction, all tables are queried in parallel for candidate results.
The selection process prioritizes predictions from tables with \emph{longer history lengths}, as these capture more detailed branch correlation patterns.

TAGE leverages tags not just for entry identification but as a fundamental mechanism for prediction selection.
When a tag mismatch occurs in a table, TAGE falls back to checking results from tables with shorter history lengths, ultimately defaulting to \texttt{T0} if no matches are found in other tables.





\subsection{Spectre Attacks}
By exploiting speculative execution driven by the BPU, researchers have discovered multiple variants of Spectre attacks~\cite{kocherSpectreAttacksExploiting2019,canellaSystematicEvaluationTransient2019,barberisBranchHistoryInjection2022,koruyehSpectreReturnsSpeculation2018,maisuradzeRet2specSpeculativeExecution2018,zhangExploringBranchPredictors2020,hofmannSpeculationFaultModeling2023,trujilloInceptionExposingNew2023,wiknerRETBLEEDArbitrarySpeculative2022} that can transform harmless memory load operations into data disclosure gadgets.
When a branch's resolution is delayed due to unresolved data dependencies, the BPU makes predictions based on patterns stored in its prediction caches.
Although mispredictions are eventually reversed through pipeline flushes, the speculative execution may perform architecturally unauthorized operations before the flush occurs. This constitutes the basis for Spectre attacks.

Beyond Spectre-v1/v2 attacks, researchers have discovered additional vulnerabilities~\cite{wiknerPhantomExploitingDecoderdetectable2023,trujilloInceptionExposingNew2023}. \emph{Straight-Line Speculation} (SLS)~\cite{armStraightlineSpeculationWhitepaper2020,wieczorkiewiczAMDBranchMispredictor2022} concerns the speculative execution of instructions immediately following 
another instruction that should change the control flow (e.g., a branch, a return, etc.). 
Branch History Injection~\cite{barberisBranchHistoryInjection2022} further exploited the interaction between PC values and BHB content in BTB indexing. By manipulating these components to generate colliding indices, attackers can force speculative execution of specific gadgets. 

Canella et al.~\cite{canellaSystematicEvaluationTransient2019} systematically categorized attack vectors based on privilege levels and execution contexts.
Their experiments revealed multiple mistraining vectors: the branch can be mistrained either {\bf\emph{in-place}} (using the vulnerable branch itself) or {\bf \emph{out-of-place}} (using a branch at a conflicting virtual address), and the mistraining can occur from either the same address space (victim process) or across different address spaces (attacker-controlled process).

Their work revealed that the effectiveness of these attacks varies significantly across platforms due to microarchitectural differences.
For instance, while out-of-place Spectre-v2 attacks were demonstrated on Intel processors, experiments on ARM's Cortex-A57 core (tested on Nvidia Jetson TX1) showed resistance to this attack vector.

\section{Microarchitectural Details of the BPU}
In this paper, we evaluate multiple processors summarized in Table \ref{tab_tested_socs}.

\begin{table}[t]
    \centering
    \begin{tabular}{lll}
    \hline
    \textbf{SoC}                                                       & \textbf{$\mu$arch} & \textbf{Linux} \\ \hline
    NXP i.MX8QM                                                       & Cortex-A72    & 5.15.71         \\ 
    BCM2712 (RaspberryPi 5) & Cortex-A76    & 6.6.63          \\ 
    Nvidia Jetson AGX Orin   & Cortex-A78AE  & 5.10.104        \\ 
    AMD Ryzen 7 7840U   & Zen4  & 6.12.20        \\ 
    Intel N100   & Gracemont  & 6.1.0 \\
    Intel Core Ultra 7 155H   & \begin{tabular}[l]{@{}l@{}}Redwood Cove\\ Crestmont\end{tabular}  & 6.8.0 \\ 
    \hline
    \end{tabular}
\caption{SoCs and Linux kernel versions tested in our paper.}
\label{tab_tested_socs}
\end{table}    

\subsection{Reference Snippet} \label{sec:snippet}
To present our following experiments, 
it is convenient to introduce a reference vulnerable code snippet in Listing~\ref{lst:snippet},
which utilizes and manipulates history-based prediction.


\begin{listing}
  \begin{minted}[
      frame=lines,
      % framesep=2mm,
      baselinestretch=0.9,
      % bgcolor=LightGray,
      fontsize=\footnotesize,
      linenos,
      highlightlines={5,6},
      highlightcolor=gray!20,
      escapeinside=||,
      ]{asm}
BH_n:           // BH[n], populate branch history
  |\textcolor{violet}{Bcond/BLR/BR}|
  // ......
  |\textcolor{violet}{Bcond/BLR/BR}|
Bx_prime:              // optional
  |\textcolor{violet}{Bcond/BLR/BR}|         // replace with padding NOPs
  B Bi_pred

\end{minted}
\vspace{-25px}
\begin{minted}[
  frame=lines,
  % framesep=2mm,
  baselinestretch=0.9,
  % bgcolor=LightGray,
  fontsize=\footnotesize,
  linenos,
  % highlightlines={6,7},
  highlightcolor=gray!20,
  escapeinside=||,
  ]{asm}
Bi_pred:
  LDR X1, target       // target=[t_safe or t_leak]
  BR X1
  
t_leak:                // alias t_primary:
  LDR X2, [X3]
  LDR X5, [X4, X2]     // refill gadget
  RET

t_safe:                // alias t_alt:
  ADD X2, X3, X4       // example benign addition
  RET
\end{minted}
\caption{Pseudocode of the reference vulnerable snippet.}
\label{lst:snippet}
\end{listing}

\textbf{\BHn} is a series of branches (of any type) that \emph{always completely} populate the BHB with a specific value \texttt{BHB(BH[n])}
\footnote{
  This notation may also be replaced with the actual BHB values, which are denoted with the notation \texttt{[\textless{}footprint\ 1\textgreater{},\ \textless{}footprint\ 2\textgreater{},\ ...]}, e.g., \texttt{[A, B, C, D]}.
}.
\textbf{\Biprobe} is an optional indirect or conditional branch that, when executed after \BHn, populates the BHB with an additional footprint.

\textbf{\Bipred} is an indirect branch with two possible targets, designated as \tsafe and \tleak. Both \BHn and \Biprobe contribute to its speculation through the combined BHB value \texttt{BHB(BH[n]Bx\_prime)}.
Upon executing \Bipred, speculative selection between these targets can be monitored through micro-architectural probes (e.g., cache hit), enabling inspection of prediction results.
  
The snippet reports a leakage gadget for target \texttt{t\_leak} and a benign operation for target \texttt{t\_safe}.
Sometimes, in the following, we just need to distinguish between these two targets, independently of the corresponding code: in these cases, the reader can ignore the leakage and benign instructions and the two targets are referred to as \texttt{t\_primary} and \texttt{t\_alt}, respectively.

Note that this is just an example of vulnerable code: in practice, many other snippet structures can be vulnerable to our attacks.

\subsection{Branch History or Path History} \label{sec:bhb-phr-exps}
We follow the method proposed by Yavarzadeh et al.~\cite{yavarzadehHalfHalfDemystifyingIntels2023}, which focused on Intel processors, to investigate BHB implementations in tested ARM and AMD processors.
Our experiments reveal that the BHBs in A76 and A78AE do not record not-taken conditional branches but can distinguish between footprints of taken branches at different addresses.
Moreover, all branch types---direct, indirect, and conditional---update a unified BHB.
These observations suggest that both models implement a PHR, rather than a canonical BHB.
ARM's official documentation~\cite{armSpectreBHBSpeculativeTarget2023} recommends BHB population loops of 24 and 32 iterations for A76 and A78AE respectively, with each iteration executing 2 branches, our experiments confirm that their PHRs can store footprints of twice these values: 48 and 64 branches, respectively.

Conversely, our experiments on A72 and Zen4 revealed a completely different implementation.
Based on our observations, we conjecture that A72 is designed with 2 separate BHBs, one canonical BHB and one PHR.
Both buffers are 8 bits in size.
The BHB holds eight 1-bit outcomes for conditional branches, while the PHR holds four 2-bit footprints obtained from the [5:4] bits of the indirect branch target.
These two buffers are updated and stored separately, then XOR'd together when read by the BPU.
AMD Zen4 also follows this design, but direct and conditional branches also update the PHR.
In the following, whenever we do not need to distinguish between these detailed  implementations, we will simply use the term BHB.

\subsection{BTB/PHT Mistraining \& Eviction} \label{sec:btb-eviction}
While untagged BTB/PHT implementations allow one branch to directly \emph{pollute (or mistrain)} the prediction of another branch due to set-index conflict, however, when a mismatching tag is detected for a committed branch, the tagging mechanism will consider it as an unrecorded one, thus will replace an existing record and resulting in \emph{eviction} of the existing record.

The effects of tagging are also evidenced in recent research.
Canella et al.'s evaluation of Spectre-v2~\cite{canellaSystematicEvaluationTransient2019} found that only Intel processors are vulnerable to the out-of-place variants, while AMD processors remain unaffected.
This observation suggests that AMD CPUs employ more comprehensive tagging strategies compared to Intel's implementations.
This is further supported by Wieczorkiewicz's work on SLS~\cite{wieczorkiewiczAMDBranchMispredictor2022}.
%
ARM processors were found to be immune to out-of-place Spectre-v2 but vulnerable to SLS for indirect branches~\cite{armStraightlineSpeculationWhitepaper2020}, suggesting that BTB eviction should also appear on these processors.
According to some papers~\cite{kiseBimodeBranchPredictor2005,changImprovingBranchPrediction1997,driesenCascadedPredictorEconomical1998}, since predictions for conditional branches may also involve BTB records, i.e., depending on branch targets, we further suspect that BTB eviction may also influence the prediction of conditional branches.

\paragraph{Mistraining.}
To verify this conjecture, we design an experiment using a Spectre-v1 snippet shown in Listing \ref{lst:btb_ev}.
We maintain one copy of the snippet as the victim while creating multiple congruent copies at 20-bit-aligned addresses (i.e., keeping lower 20 bits identical to ones in the address of the victim) to serve as mistrain snippets.



\begin{listing}
\begin{minted}[
    frame=lines,
    % framesep=2mm,
    baselinestretch=0.9,
    % bgcolor=LightGray,
    fontsize=\footnotesize,
    linenos,
    highlightlines={4}
    ]{c}
void test_eviction(bool *flag, char *dc_signal) {
    populate_bh(); // or replace with padding NOPs
    if ( *flag == 0 )
        char junk = *dc_signal;
}
\end{minted}
\caption{Snippet to test BTB/PHT mistraining.}
\label{lst:btb_ev}
\end{listing}

\begin{figure}[t]
    \centering
    \includegraphics[width=\linewidth]{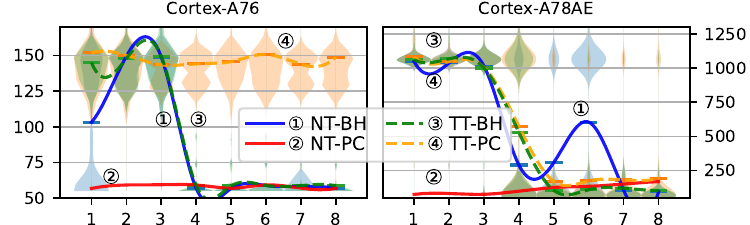}
    \caption{
    Cache access latency (ns) as a function of the number of mistrain snippets (x-axis) after out-of-place training of victim if-load snippet. Lower latency indicates not-taken speculative execution of the victim branch. Victim branch initially trained to not-taken (NT) or taken (TT), with out-of-place training using matching (BH) or different branch histories (PC). Results averaged over 12,800 tests per configuration.
    }
    \label{fig_out-of-place_training}
\end{figure}

The experiment begins with establishing a branch highly biased to \emph{not-taken} by repeatedly executing (e.g., 32 times) the victim snippet with \texttt{flag=0} (NT-* in Fig.~\ref{fig_out-of-place_training}).
Following this initialization, out-of-place training attempts to mistrain the branch to \emph{taken} by executing multiple mistrain snippets with \texttt{flag=1}.
To observe the prediction outcome through data cache signals, the victim snippet runs again with \texttt{flag=1} set.
We also further evaluate the opposite case in which the victim branch is initially trained as taken (\texttt{flag=1}) (TT-* in Fig.~\ref{fig_out-of-place_training}).

\paragraph{Varying the number of mistrain snippets.}

We first conducted the test on A76 and A78AE. Results are reported in Fig.~\ref{fig_out-of-place_training}.
Our tests start with no mistrain snippets as a baseline, then increase the number from one to eight.
The baseline test without mistrain snippets successfully establishes the biased prediction.
When initially training the victim branch as not-taken, if using 1 to 3 mistrain snippets, the cache signal disappears, indicating the branch is successfully mistrained to taken through out-of-place Spectre-v1 attack (NT-BH in Fig.~\ref{fig_out-of-place_training}).
Interestingly, starting from 4 branches, the cache signal reappears, suggesting an unexpected change in BTB and PHT behavior.Even when initially training the victim branch as taken (\texttt{flag=1}), exceeding this threshold of congruent branches causes the prediction to revert to not-taken (TT-BH in the figure).
Similar patterns emerge on A78AE, where out-of-place Spectre-v1 attacks only succeed with one to four mistrain snippets.
Furthermore, on A72, just two mistraining snippets were sufficient to trigger this inversion phenomenon.

We also performed the same experiment on x86 processors.
While we were able to perform regular out-of-place mistraining for conditional branches on all tested processors, similar reverting behavior appeared on AMD Zen4 when executing 16 mistrain snippets containing conditional branches jumping across a 4K-aligned boundary.
However, this reverting behavior was not observed on Intel processors, suggesting a different implementation of tagging mechanisms.

This experiment demonstrates that, in tested ARM and AMD processors, while saturation counters can be shared among multiple branches and lead to out-of-place mistraining, exceeding a threshold number of congruent branches accessing the same branch prediction entry triggers an eviction-like behavior for conditional branches.
This may not match the statement of ARM claiming only unconditional branches vulnerable to SLS~\cite{armStraightlineSpeculationWhitepaper2020}.
We tentatively attribute this phenomenon to \emph{BTB/PHR eviction} mechanisms.

Mistraining and eviction with different branch histories were also tested: details are available in Appendix~\ref{sec:btb-ev-diff-hist}.

\section{Threat Model}
We consider a data disclosure threat model where the attacker possesses knowledge of targeted hardware and can identify or inject vulnerable execution patterns in the victim system. The unprivileged or privileged attacker can execute the corresponding vulnerable code snippets. The target system has no software vulnerabilities. All recommended Spectre mitigations are enabled with recommended configurations unless explicitly noted as being disabled for specific experiments.

\section{Exploitation 1: Bias-Free Branch Prediction}


During our tests on the BHB of A72, we observe that an indirect branch consistently jumping to the same target from program initialization (e.g., a fixed function call in C code compiled as an indirect branch) never updates the path history.
%
This unusual behavior of A72 suggests an undocumented BHB and PHR update mechanism that selectively records branch footprints based on certain conditions.
We attribute this behavior to \textbf{Bias-Free Branch Prediction}~\cite{gopeBiasFreeBranchPredictor2014,al-otoomDetectingFilteringBiased2014}, and present primitives exploiting this undocumented behavior.

\subsection{Filtering Biased Branches from BHB} \label{sec:bias-free}
In history-based branch prediction, the fixed size of the BHB imposes a limitation on its capacity to store control flow information.
To maximize prediction accuracy, the BPU must effectively eliminate irrelevant data from the control flow, ensuring the BHB retains older yet meaningful footprints within the limited storage budget.

One significant source of irrelevant information is \emph{biased branches}.
These are branches that consistently produce the same outcome throughout a program's execution.
A biased \emph{conditional} branch always maintains the same taken or not-taken status, while a biased \emph{indirect} branch consistently jumps to the same target address.
Since biased branches neither affect future control flow nor depend on earlier branches, their footprints do not contribute to meaningful history for branch prediction.
Instead, they would occupy space in the buffer, reducing its capacity to capture correlations from earlier, more informative branches, thereby degrading prediction accuracy.

The \textbf{Bias-Free Branch Predictor}~\cite{gopeBiasFreeBranchPredictor2014,al-otoomDetectingFilteringBiased2014} was introduced building on this intuition.
Although the proposed implementations differ in their details, their key innovation lies in optimizing the BHB update mechanism by assessing the bias status of a branch before adding its footprint to the BHB.
This ensures that only footprints from non-biased branches are recorded, effectively preserving space for older branch histories. 

\begin{figure}
\centering
\includegraphics[width=0.72\linewidth]{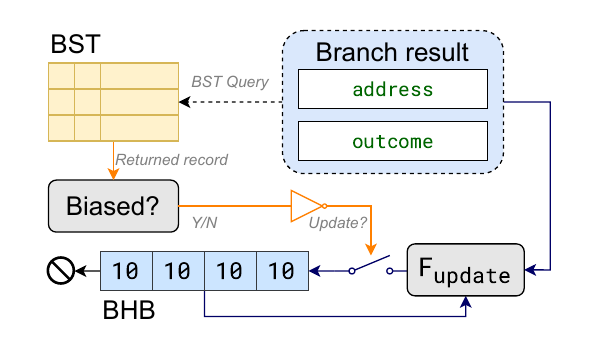}
\caption{BHB update process in bias-free branch prediction. When a branch resolves, the BPU determines the branch's bias status using the process described in Algorithm~\ref{alg_bias_determination}, excluding biased branches from the updating process.}
\label{bias_free}
\end{figure}

This method incorporates a logical \textbf{Branch Status Table (BST)} alongside the conventional buffers and registers used in history-based branch predictors.
As shown in Figure \ref{bias_free}, the BST plays a critical role in the branch history update process by tracking the bias status of executed branches.
Each \emph{BST entry} contains two fields: \emph{last branch outcome} and \emph{bias status}, which together represent the status of branches. 

The interested reader can refer to Appendix~\ref{app:bias-algo} for the update algorithm of BST entries.
Most importantly, note that a branch that has never been seen before is always considered biased, as there is no evidence to suggest otherwise.
This status only changes when the branch produces a secondary outcome that differs from the previously recorded one, at which point the relevant BST record is updated.

\paragraph{Experimental validation.}
To investigate this feature's behavior, we employed the reference snippet in Listing~\ref{lst:snippet}.
In our experimental setup, \BHn functions as a chain of indirect branches and conditional branches designed to fully populate the BHB and flush information from the previous context.
The jump targets of indirect branches are carefully selected to chain them together while ensuring each branch consistently jumps to the same target address.

Based on this controlled BHB value, the prediction of \Bipred should theoretically be manipulable through the outcome of \Biprobe.
On most processors we evaluated, we consistently observed that the prediction of \Bipred is indeed influenced by the outcome of \Biprobe, confirming the expected behavior of standard history-based prediction.

However, during our tests on the A72's BHB, we observed that the processor consistently yielding predictions matching the architectural target.
This observation suggests that such a chain of branches fails to fully update the path history, leaving residual hints from earlier contexts within the BHB.
By methodically chaining the indirect branches and directing them to different targets prior to initiating our tests, we were able to achieve the aforementioned control until process termination using the same branch chain.
Interestingly, interference from other processes sharing the same processor core could subsequently disrupt this control mechanism.

These distinctive behaviors strongly indicate that the A72 may have implemented a bias-free branch predictor that utilizes a globally shared Branch Status Table.

\subsection{BST Eviction} \label{bst-eviction}

As Gope and Lipasti~\cite{gopeBiasFreeBranchPredictor2014} suggested, the BST should be implemented as a fully-associative table indexed by the lower bits of branch addresses.
To enhance isolation among different contexts, the BST may include an additional \emph{tag} field for each entry, which is generated using a different hash function from the one used for indices.
By verifying the \emph{tag} value upon a query, it prevents the retrieval of records assigned to a different branch, thus mitigating potential value injection.

However, similar to other caches that use tags for isolation, in the context of the BST, \emph{eviction} occurs when a branch attempts to acquire the slot that is already occupied by a victim.
While the new branch will replace the existing entry with its own data, the record associated with the victim branch is removed from the BST.
When the victim branch is executed again after eviction, \emph{the victim branch will be classified as biased}, regardless of its previous behavior before eviction.

\paragraph{Observing BST eviction.} \label{observing_bst_e}

Based on the interaction between the BHB and BST discussed above, we note  that BST eviction can be monitored by observing its side effects on history-based branch prediction.
Building on this premise, we demonstrate how BST eviction can be observed by \emph{monitoring mis-speculations triggered by inaccurate branch histories}.




Consider the code snippet of Sec.~\ref{sec:snippet}. We train the BPU to predict \Bipred under two distinct execution flows: 
\begin{itemize}
\item \textbf{\flowa}: it invokes \BHn, skips the optional \Biprobe, and then causes \Bipred to jump to \tprimary; and
\item  \textbf{\flowb}: it invokes \BHn and \Biprobe in sequence, then causes \Bipred to jump to \talt.
\end{itemize}
  

Due to the presence of \Biprobe, \flowb generates a BHB value for \Bipred that differs from the one generated by \flowa.
When alternatively executing \flowa and \flowb under normal conditions, the BPU should be able to differentiate these two flows and make accurate predictions based on the following BTB entries\footnote{Notation \texttt{(x,\ BHB(y))\ →\ z}  represents a BTB entry, indicating that branch \texttt{x} is predicted to target \texttt{z} when the branch history is \texttt{BHB(y)}.}:
%
%
\begin{equation}
\begin{aligned}
  &\mathbb{F}_A: (\texttt{Bi\_pred}, ~\texttt{BHB(BH[n])}) \rightarrow \texttt{t\_primary}, \\
  &\mathbb{F}_B: (\texttt{Bi\_pred}, ~\texttt{BHB(BH[n]+Bi\_probe)}) \rightarrow \texttt{t\_alt}.
\end{aligned}
\end{equation}

We now introduce \textbf{\Bxevict}, a branch that contends for the same BST entry as \Biprobe (see Listing~\ref{lst:snippet}) through index aliasing, though it exists outside our reference snippet.
\Bxevict causes the eviction of the BST entry upon its execution, altering the branch prediction status of \Biprobe.
To demonstrate the effects of BST eviction, we can hence invoke \texttt{Bx\_evict} in the middle of the two flows \flowa and \flowb. 
On the execution of \flowb, this eviction will force \Biprobe to be classified as biased.
The footprint of \Biprobe will hence be omitted, and the BHB will be populated solely based on the footprints of \BHn as for \flowa.

Consequently, despite executing \flowb, the BTB entry for \flowa will be used for predicting \texttt{Bi\_pred}, leading to \texttt{t\_primary} being mis-speculated in the mismatching context of \flowb.


\paragraph{BST on Cortex-A72.}

We implemented the reference snippet from Listing~\ref{lst:snippet} as a userspace program to test this behavior on A72 using the NXP i.MX8QM chip.
In this program, we define \Bxevict as an always-taken conditional branch and copy it to an address that shares at least 16 lower bits with with \Biprobe at program initialization.
\tprimary triggers a cache fetch to a designated probe address, while \talt performs no observable operations.

The mis-speculation was observed with 100\% success rate, demonstrating the presence of a Bias-Free Branch Predictor in the ARM A72 CPU. We also found that the logical BST is implemented as a 4096-entry table indexed by a 12-bit value derived from bits [15:4] of an instruction's virtual address.
The successful eviction with a single branch indicates that the table is full-associative and tagged.
Since \BHn is not called before \Bxevict, the high success rate of mis-speculation indicates that branch history is not involved in the indexing process.
Our experiments further revealed other possible sources of BST eviction, as detailed in Table~\ref{bst_evict}, which we confirmed by testing different branch types for \Bxevict.

\begin{table}
\centering
\begin{tabular}[]{@{}lll@{}}
\hline
\textbf{Type} & \textbf{Mnemonics} & \textbf{Evict?} \\
\hline
Indirect & \texttt{BR}, \texttt{BLR} & Yes \\
Conditional & \texttt{B.cond}, \texttt{TB(N)Z}, \texttt{CB(N)Z} &
\emph{When taken} \\
Unconditional & \texttt{B}, \texttt{BL} & No \\
Return (indirect) & \texttt{RET} & No \\
Other & \texttt{SVC} & No \\
\hline
\end{tabular}
\caption{Operations triggering BST evictions on Cortex-A72.}
\label{bst_evict}
\end{table}

Additionally, we find that this behavior differs for conditional branches: details are available in Appendix~\ref{sec:bhb-zen4}.

\paragraph{Cross-context eviction.}
We further test if this primitive can bypass process isolation, privilege levels, and Spectre mitigations.
We implement the reference snippet in a custom system call handler in Linux to check whether a userspace \Bxevict can affect branch prediction in kernel space.
While we can train the BPU to distinguish \flowa and \flowb through the \texttt{syscall()} interface, \Bxevict can still induce mis-speculation in \flowb and leave an observable data cache signal.

This result reveals gaps in the isolation and sanitization of BST by current Spectre software-based mitigations.
The ARM-proposed Spectre-BHB mitigation~\cite{armSpectreBHBSpeculativeTarget2023} can effectively prevent explicit BHB value manipulation through population, but our kernel-mode proof of concept demonstrates a lack of sanitization of update policy after this barrier, creating a residual attack surface for crafting BHB values.

In the second experiment, we implement the reference snippet and \Bxevict as separate userspace programs.
The victim program contains the reference snippet, while \Bxevict is placed at the BST-aliasing address and runs in an infinite loop.
Both programs run concurrently on the same core, with the victim program actively yielding CPU time to allow \Bxevict execution.
However, when the victim program regains the processor, we can neither observe the data cache signal nor detect the previously established entries for \flowa and \flowb.

The Spectre-v2 mitigation on A72 employs a BPU flush implemented in the ARM Trusted Firmware to invalidate all BPU information~\cite{armTrustedFirmwareA21202018}.
While this reset can effectively sanitize all prediction state, we find that Linux applies it only to userspace context switches, similar to the restricted scope of IBPB on x86 processors.~\cite{intelIndirectBranchPredictor2018,LinuxKernelDocSpectre}
With this mitigation disabled, we successfully observed mis-speculation caused by eviction from another userspace process.

Recent ARM processors implement a hardware-based isolation feature \texttt{FEAT\_CSV2}~\cite{armArmArchitectureReference2024} that adds context-dependent values to BTB tags. However, although our tested processor revision (r0p2) does not support this feature, ARM's feedback confirms that such eviction remains unrestricted by this feature.



\subsection{Attack Flow \#1: BiasScope} \label{biasscope}
Building on BST eviction, this section presents \textbf{BiasScope}, a side-channel attack that leverages BST features to leak the outcome of branches, even in other exception levels.

In spirit, BiasScope offers capabilities similar to those of \emph{BranchScope}~\cite{evtyushkinBranchScopeNewSideChannel2018}, but leveraging a different attack vector. BranchScope is a side-channel attack that extracts coarse-grained control-flow information by analyzing saturation counter values in the \emph{Pattern History Table (PHT)} of modern branch predictors. Conversely, our BiasScope exploits BST eviction to monitor the execution flow of a victim program, resulting from a taken branch that evicts an existing BST entry with a shared aliasing index.

The core concept of BiasScope builds upon the primitive described in Section~\ref{observing_bst_e}.
BiasScope does not perform the eviction between \flowa and \flowb.
Instead, it initializes and maintains the \emph{non-biased} record for \Biprobe, and yields the processor to the victim process.
By alternating the execution of \flowa, \flowb, and the victim process, the attacker can detect whether a secret-dependent branch in the victim context triggered a BST eviction, effectively repurposing the BST itself as a side channel to extract information from other processes.

\begin{figure}[t]
  \centering
  \includegraphics[width=0.9\linewidth]{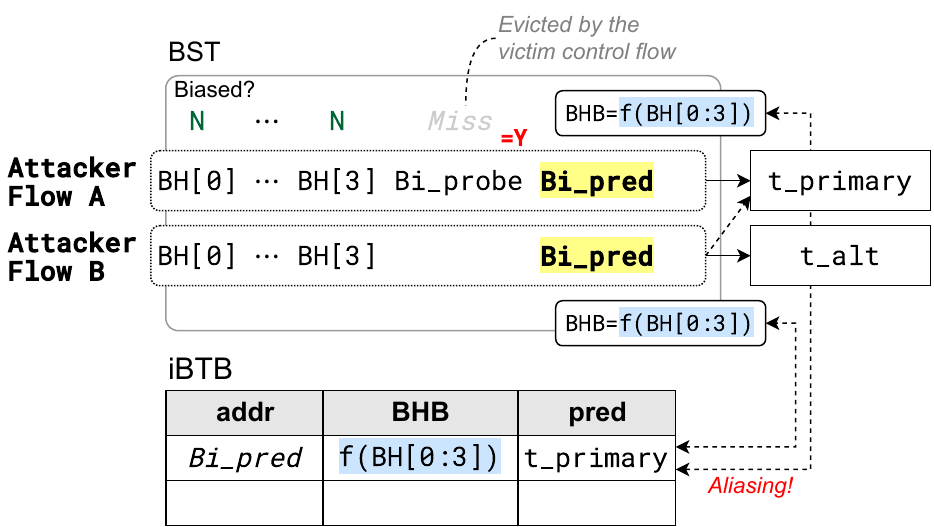}
  \caption{Monitoring a victim branch using BiasScope.}
  \label{fig_biasscope}
\end{figure}  

The attack flow is depicted in Figure~\ref{fig_biasscope}.
This BST side-channel involves a secret-dependent \textbf{sender} branch in the victim context and a \textbf{receiver} snippet controlled by the attacker.
The sender branch corresponds to \Bxevict and \Biprobe must be selected by the attacker so that the two branches share the same BST entry (e.g., same address bits [15:4] on Cortex-A72).
According to the BST eviction behavior we found, it can be a \emph{bare} conditional branch that can be monitored directly, or an indirect branch \emph{nested} within a conditional block, thereby exposing the execution status of the preceding conditional branch.
The receiver leverages flows \flowa and \flowb introduced in the previous section, utilizing the history-related components \BHn, \texttt{Bi\_pred}, and \Biprobe to determine whether the sender branch was taken.


The BiasScope attack proceeds as follows:

\begin{enumerate}
\item
  \textbf{Preparation}: The attacker first forces \Biprobe to alternate between two legit targets to establish its \emph{non-biased} record in the BST. Additionally, the attacker ensures that the branches in \BHn are recorded as \emph{non-biased} to fully populate the BHB.
\item

  \textbf{Victim Execution}: The attacker yields control of the CPU, allowing the secret-dependent sender branch to execute in the victim context.
\item

  \textbf{Observation}: After regaining CPU control, the attacker alternates between executing \flowa and \flowb to verify the presence of the BST entry of \Biprobe.
\end{enumerate}

If the sender branch was taken during this period, the \emph{non-biased} status of \Biprobe will be lost.
Consequently, the BPU will classify \Biprobe as a biased branch, omitting its footprint in \flowb and causing the BHB value to match that of \flowa.
In this scenario, \texttt{t\_primary} is mis-speculated in \flowb, which can be detected using a micro-architectural probe (e.g., cache hit).
This allows the attacker to leak the state of the victim branch, possibly revealing secrets in the victim context.

\begin{figure}[t]
  \centering
  \includegraphics[width=\linewidth]{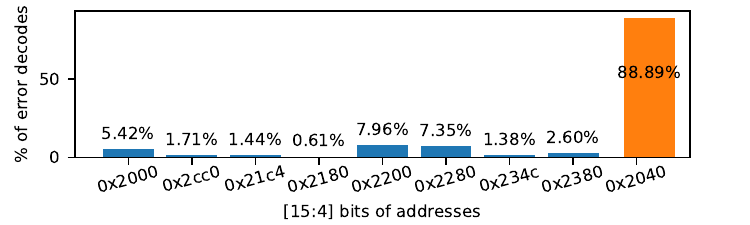}
  \caption{Error rate of BST side channels under intra-process BiasScope with branch addresses with different [15:4] bits.}
  \label{fig:bs_snr}
\end{figure}

\paragraph{Evaluation.}
We first evaluated whether we could leak the taken status of an injected kernel branch with \emph{all default mitigations enabled}.
While controlling its conditional direction from userspace, our results demonstrate that this vector can accurately detect taken events of conditional branches executing in kernel space.
We further evaluated the performance of this BST side-channel using both sender and receiver running in userspace, disabling the Spectre-v2 mitigation (discussed in Section~\ref{sec:bias-free}) for the sole purpose of this experiment.
The sender encodes an 8-bit secret using eight independent conditional branches, with each branch controlled by one bit of the secret byte. In each iteration, the receiver yields the core to the sender by sleeping briefly, allowing the sender to encode a secret byte into the BST side channel. When the receiver resumes execution, it attempts to decode all eight bits.
The error rates in decoding each bit are illustrated in Fig.~\ref{fig:bs_snr}.
Our experiments demonstrate a high signal-to-noise ratio side channel.
However, we also observe that some branch addresses (e.g., with bits [15:4] = \texttt{0x2080}) become completely jammed for certain periods, suggesting interference from other code snippets sharing the same processor core.

BiasScope converts the presence of a BST entry into observable branch latency.
To effectively perform the attack, the attacker must have a detailed understanding of the target CPU's branch latency characteristics to decode the measured branch latency.
Furthermore, since the BST can track multiple branches simultaneously, BiasScope can monitor several non-aliasing branches concurrently, improving the granularity of observations on the victim's execution flow.

\subsection{Attack Flow \#2: Spectre-BSE} \label{spec-bse}
While BiasScope demonstrated how data leakage can be facilitated by observing BST eviction caused by the victim, we now exploit this behavior in the opposite direction, triggering malicious mis-speculations in the victim context.
We introduce \textbf{Spectre-BSE (Branch Status Eviction)}, a novel \emph{target reuse attack} primitive that exploits BST evictions to facilitate BHB aliasing and trigger malformed speculative executions.

As we demonstrated before, since the bias-free mechanism has an obvious influence on the BHB updating policy, an attacker may manipulate the generation of BHB value by controlling the presence of relevant BST records.
This may cause unexpected BHB values, which can be further exploited to induce BTB index aliasing, reaching a similar result to \emph{Branch History Injection} (Spectre-BHB)~\cite{barberisBranchHistoryInjection2022}.
In this section, we demonstrate how BST eviction can manipulate branch prediction and trigger secret data disclosure.

Spectre-BHB alters the BTB query and selection behavior by manipulating the BHB values with footprints from attacker-controlled branches, unlike Spectre-v2, which directly injects a BTB entry into the target entry.
In history-based BPUs, the BTB index function typically incorporates both the branch address and the BHB value.
This dependency can be exploited by crafting malicious BHB values, leading to confusion between two different branches.
This behavior enables a ``Target Reuse Attack'', facilitating implicit \emph{out-of-place} branch target injection and bypassing existing Spectre-v2 mitigations.

However, Barberis et al.~\cite{barberisBranchHistoryInjection2022} stated that they were unable to reproduce out-of-place Spectre-BHB attacks on ARM devices.
Considering that our attack Spectre-BSE ultimately relies on the same BTB indexing mechanism as Spectre-BHB, we currently limit our demonstration to in-place branch target training, while demonstrating an inherent out-of-place vector for manipulating BHB updates on the Cortex-A72 processor.
It is important to note that since neither Barberis et al.~\cite{barberisBranchHistoryInjection2022} nor ARM~\cite{armSpectreBHBSpeculativeTarget2023} has completely ruled out the possibility of out-of-place BHI attacks on ARM processors, we conjecture that it remains feasible to conduct a \emph{fully} out-of-place target reuse attack using our Spectre-BSE.

Consider the reference snippet of Sec.~\ref{sec:snippet} without \Biprobe and the following execution flows: 
\begin{itemize}
\item \flowa: it invokes \texttt{BH[n]} then \texttt{Bi\_pred} jumps to a disclosure gadget \textbf{\texttt{t\_leak}}; and
\item \flowb: it invokes \texttt{BH[n]}, sets a secret-related context (e.g., in registers), and eventually jumps to \textbf{\texttt{t\_safe}}, a benign target that poses no security risks.
\end{itemize}

While branches in \texttt{BH[n]} may, e.g., validate the passed parameters to prevent micro-architectural illegal memory loads, they should also create distinct BHB values, resulting in two BTB entries corresponding to \flowa and \flowb, respectively.
Similar to typical Spectre attacks, triggering the mis-speculation of \texttt{t\_leak} in a mismatching context, i.e., \flowb, allows the attacker to induce data disclosure from a secret context.

BST eviction plays a critical role in this attack by \emph{forcing a typically non-biased branch to be classified as biased}, thereby generating an unexpected BHB value in the BTB query or update process.
When a branch is executed after the corresponding BST entry is evicted, the BHB is not updated, causing one oldest footprint to remain inside the BHB, leading to an unexpected BHB value in subsequent control flows.
In some occasions, this may make the BHB value used to predict \texttt{Bi\_pred} alias with a mismatching execution flow, causing the wrong record to be used in the prediction.

The Spectre-BSE attack hence proceeds as follows:
\begin{enumerate}
\item \textbf{Preparation:} 
Exploitability hinges on the BHB footprint of \texttt{BH[n]} in both flows \flowa and \flowb.
The attacker identifies flows \flowa and \flowb such that \texttt{BH[n]} generates two footprint sequences: \texttt{BHB(BH[n]$\mid$\flowa)} and \texttt{BHB(BH[n]$\mid$\flowb)}.
Exploitation is possible if, by excluding a subset \(\mathcal{B}_{ev} \subseteq \texttt{BH[n]}\) from \flowb, BHB aliasing occurs, i.e.,
\begin{equation}
  \begin{aligned}
  \texttt{BHB(BH[n]$\mid$\flowa)}&=\texttt{BHB(BH[n]-$\mathcal{B}_{ev}$$\mid$\flowb)}.
  \end{aligned}
\end{equation}
The attacker then invokes \flowa to initialize a BTB entry.
\item \textbf{Eviction:} The attacker performs a targeted BST eviction on $\mathcal{B}_{ev}$ using another branch that contend for the same BST entry due to aliasing.
\item \textbf{Leakage:} The attacker invokes \flowb, inducing mispeculation towards \texttt{t\_leak} while retaining a secret-related context.
\end{enumerate}

Compared to Spectre-BHB~\cite{barberisBranchHistoryInjection2022}, Spectre-BSE manipulates BHB values through BST eviction rather than directly injecting attacker-controlled branches.
Furthermore, it does not require a short execution path between the snippet entry and the victim branch, allowing for more flexibility in attack scenarios and significantly broadening the potential attack
surfaces.

Suppose \texttt{BH[n]} comprises five indirect branches, labeled as \texttt{BH[0]} through \texttt{BH[4]}.
All these branches are initially non-biased, and their bias statuses are trained and stored in the BST before the attack.
Under unconstrained BHB budget capacity, the execution of this branch sequence would yield the following BHB values for the two flows:
%
\textbf{(i)}  \texttt{BHB(BH[n]$\mid$\flowa)}=\texttt{[A,B,C,D,E]};
\textbf{(ii)}  \texttt{BHB(BH[n]$\mid$\flowb)}=\texttt{[B,C,D,E,F]},
where capital letters denote some example BHB values.
As the budget capacity of BHB is limited in practice, let us assume the BHB retains only the four most recent footprints (as, for instance, we found happening in Cortex-A72).

To setup the attack, we first invoke \flowa to initialize a BTB entry.
This results in the following BTB entry:
\begin{equation}\label{eq:BSE-BST-Fa}
  \begin{aligned}
    &\mathbb{F}_A: (\texttt{Bi\_pred}, ~\texttt{[B,C,D,E]}) \rightarrow \texttt{t\_leak}.
  \end{aligned}
\end{equation}

\begin{figure}
  \centering
  \includegraphics[width=\linewidth]{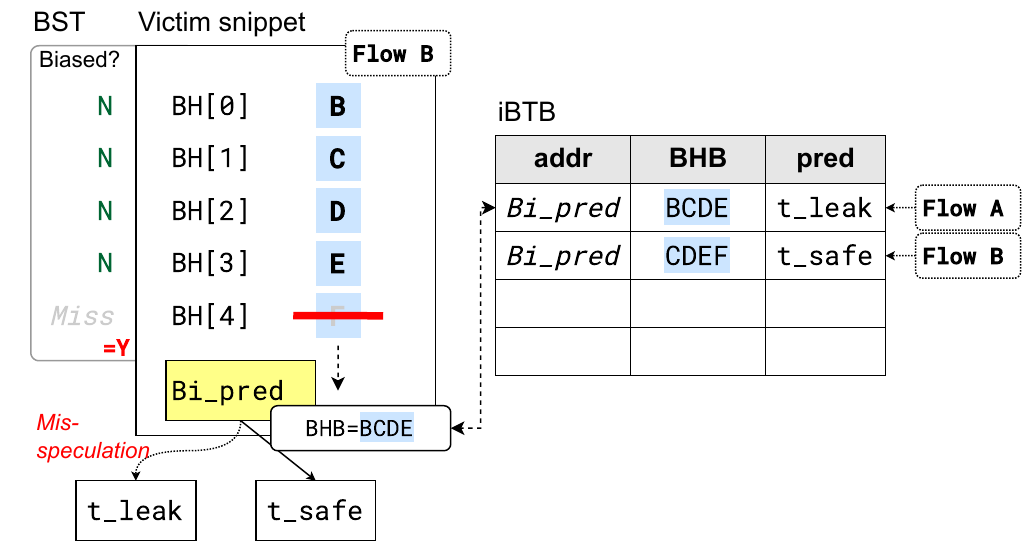}
  \caption{BHB aliasing in Spectre-BSE.}
  \label{fig_spec_bse}
\end{figure}

At this stage, the environment is almost prepared, and the attacker is ready to proceed with the malicious actions.
The attacker performs a targeted BST eviction on $\mathcal{B}_{ev}=\texttt{BH[4]}$, then invokes the vulnerable code snippet with \flowb{}.

As depicted in Fig.~\ref{fig_spec_bse}, due to eviction, the BST query for \texttt{BH[4]} results in a miss, causing its footprint to be omitted during BHB updates.
Hence, differently from the the nominal case in which \(\texttt{BHB(BH[n]$\mid$\flowb)}=\texttt{[C,D,E,F]}\), upon \flowb's execution the BHB value will be updated to \(\texttt{BHB(BH[n]-$\mathcal{B}_{ev}$$\mid$\flowb)}=\texttt{[B,C,D,E]}\) instead, with the footprints of \texttt{BH[0:3]} remaining inside the buffer and aliasing with the value associated with \flowa (see Eq.~\eqref{eq:BSE-BST-Fa}).
This maliciously-constructed BHB value forces the BPU to speculate \texttt{t\_leak} instead of \texttt{t\_safe} for \texttt{Bi\_pred}, potentially exposing sensitive data during transient execution.

\paragraph{Evaluation.}
We evaluate our exploit using a userspace branch to evict victim branches in both the same process and a \texttt{syscall()} handler with \emph{all default mitigations enabled}. When the eviction branch is placed at a 32-byte aligned address sharing bits [15:4] with the victim, we observe cache hits from the victim flow with 99.9\% success rate in both intra-process and cross-privilege attacks, consistently with our previous findings.

\section{Exploitation 2: Branch History Speculation}
This section presents Spectre-BHS.
Before proceeding, it is necessary to introduce Branch History Speculation.
\subsection{Early BHB Updates}
Due to the significant speed disparity between memory access and pipeline execution in modern processors, branch resolution can be delayed by up to hundreds of cycles.
Speculative execution enables the CPU frontend continue filling the pipeline by speculatively executing instructions during this period, potentially issuing hundreds of uncommitted instructions within the speculation window.
While this technique is essential for pipeline efficiency, it introduces new challenges when \emph{branches appear within the speculation window}.

To maintain backend utilization and avoid pipeline stalls, the frontend must predict and execute additional branches encountered in the speculative execution path until backend resources are exhausted, 
rather than halting speculation upon encountering new branches.

Although history-based branch prediction is widely adopted for this purpose, predicting a branch within the speculation window poses a unique challenge: since preceding branches may remain uncommitted and the execution path is still speculative, constructing an accurate branch history for the current prediction becomes difficult. 
%
However, if the BHB updates only upon branch resolution, these predictions, made inside speculation windows, will rely on an outdated branch history, if any exists.

To address this limitation, it becomes essential to \emph{update the BHB speculatively based on predictions}, even before branch outcomes are confirmed, rather than waiting for the commit stage which would significantly delay predictions.
Modern BPUs introduce \textbf{Branch History Speculation (BHS)}~\cite{haoEffectSpeculativelyUpdating1994} through various rollback mechanisms~\cite{golanderCheckpointAllocationRelease2009, skadron2000speculative, seznecAnalysisOGEometricHistory2005}.
This approach allows speculatively-predicted branch outcomes to immediately update the global branch history, either in the main BHB or a dedicated speculative history buffer.

Since speculative predictions become immediately visible to subsequent branches in the speculation window, the BHB remains up-to-date for further predictions.
This allows the pipeline to follow previously-learned execution paths, improving efficiency even when data dependencies remain unresolved.




\subsection{Attack Flow \#3a: Spectre-BHS} \label{spec-bhs-e}

This mechanism further implies that a branch predicted in the speculation window may be influenced by the outcomes of earlier branches that are also speculated but not yet resolved.
To systematically investigate how unretired instructions impact early BHB/PHR updates, we further extend the experiment in Section~\ref{sec:btb-eviction} in combine with the reference snippet in Section~\ref{sec:snippet}.
Through this analysis, we introduce Spectre-BHS, a novel variant of Spectre attack that manipulating the BHB updating mechanism through BTB/PHT mistraining.

\paragraph{Cascaded mis-speculation.}

In our experimental setup, we configure \Biprobe as a \emph{conditional} branch dependent on a variable \texttt{flag}.
While \BHn populates the BHB with a predetermined path, the prediction of \Bipred is principally determined by the outcome of \Biprobe.

We define two flows, \flowa and \flowb, similar to those in Section~\ref{bst-eviction}, which create distinguishable BTB/PHT entries for \Bipred based on \Biprobe's footprint:

\begin{itemize}
\item \flowa: \Biprobe is \emph{not taken} then \Bipred jumps to \tleak; and
\item \flowb: \Biprobe is \emph{taken} then \Bipred jumps to \tsafe.
\end{itemize}

Following the methodology detailed in Section~\ref{sec:btb-eviction}, we further evaluated whether we could manipulate the prediction of \Bipred by controlling the outcome of \Biprobe.
We started from single mistraining branch 
The results demonstrate that all processor cores we tested consistently yielded the expected prediction of \Bipred based on the outcome of \Biprobe, providing strong evidence that the BHB is indeed updated speculatively across all evaluated microarchitectures.
This confirms that history-based branch predictions are consistently made using the most recent branch histories, even when those histories include speculative branches.

To validate this hypothesis, we constructed multiple address-congruent mistraining snippets, each containing the identical \BHn sequence and a mistraining conditional branch \texttt{Bc\_mt} strategically placed at addresses that conflict with \Biprobe (i.e., sharing the same lower address bits).
This experimental configuration enables us to systematically mistrain \Biprobe's prediction by executing these conflicting branches while simultaneously establishing a conflicting history pattern in the PHR.


Out test works as follows:
\begin{enumerate}
\item 

\textbf{Preparation:} We repeatedly invoke \flowa and \flowb to train the BPU to recognize both flows, letting \Bcfp to be recorded as taken in \flowb.
\item 

\textbf{Mistraining:}
Then, we invoke all mistraining snippets, with the mistraining branch  to influence \Biprobe's prediction record.

\item 
\textbf{Mis-speculation:} To ensure a large speculation window that involves both \Biprobe and \Bipred, we evict from cache variable \texttt{flag} and the pointer variable 
used by indirect branch \Bipred.
Finally, we execute \flowb and monitor for the presence of the data cache signal left by \tleak.
\end{enumerate}

Similar to the experiment in Section~\ref{sec:btb-eviction}, we initiated our testing with a single mistraining snippet.
Through manipulation of the direction of \texttt{Bc\_mt} with just one mistraining snippet, we could control the speculated target of \Bipred with approximately 100\% success rate on all evaluated platforms.
This result demonstrates that speculated branch outcomes can indeed update branch history and influence subsequent speculations within the same speculation window.

\paragraph{BTB/PHT eviction in PHR.}
Our observations in Sec.~\ref{sec:bhb-phr-exps} show that Cortex-A76 and A78AE employ path history (see also Sec.~\ref{sec:branch-pred}),
where \Biprobe updates the history in PHR only when taken.
Based on this architectural insight, we hypothesize that when BTB/PHT eviction forces the BPU to ``forget'' a branch, such a not-yet-recorded branch will not update the BHB before it is detected, causing it to be \emph{implicitly misinterpreted as not-taken} in the path history. 
This creates confusion between \flowa and \flowb during speculation so that \(\texttt{PHR(BH[n]$\mid$\flowa{})}=\texttt{PHR(BH[n]-$\mathcal{B}_{ev}$$\mid$\flowb{})}\).
This could induce mis-speculation to \tleak in the context of \flowb, which can be exploited to achieve data disclosure like existing Spectre attack vectors.

On A76 and A78AE, as the number of \emph{taken} mistraining branches increased. we observed the processor began selecting the path with \Biprobe as not taken, leading to speculating \tleak.
This behavior closely parallels the phenomenon documented in Section~\ref{sec:btb-eviction}.
In our userspace testing of 12,800 trials, we successfully achieved mis-speculation through BTB eviction with success rates of 99.84\% and 99.29\% on A76 and A78AE respectively.
We extended this experiment with kernel-space victims on A76, achieving a comparable success rate of 99.33\%.
This result confirms that PHR is prone to transient confusion about the branch outcomes based on incomplete and unconfirmed branch history. 
Additionally, as we previously excluded the possibility of inducing BTB/PHT eviction in Intel processors, we were consequently unable to replicate this specific eviction-based mis-speculation vector across the Intel architecture family.

\paragraph{BTB/PHT eviction in BHB+PHR.}
Having established different BHB implementations in Cortex-A72 and AMD Zen4, we could not reproduce the eviction-induced behavior on these processors as we discussed above.
This indicates fundamental architectural differences in branch history management, which we attribute to their hybrid BHB+PHR implementations.
For a comprehensive analysis of these architectural differences and potential attack surface, see Appendix \ref{sec:bhb-zen4}.

\paragraph{Speculation window.}

\begin{figure}
  \centering
  \includegraphics[width=\linewidth]{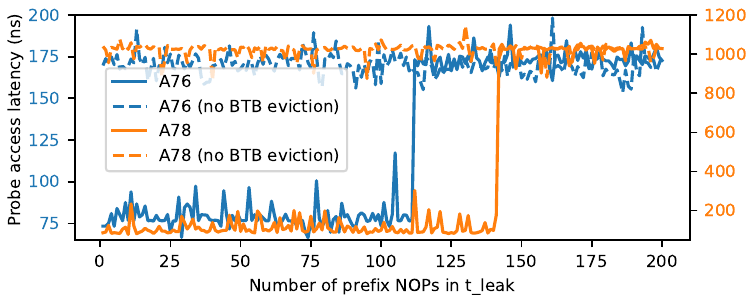}
  \caption{Access latency of data cache probe with varying numbers of prefix \texttt{NOP} instructions in \tleak{}. Lower latency indicates successful speculative execution of data load. Results averaged over 1,280 tests per configuration, shown with and without BTB evictions.}
  \label{fig:spec_bhs_window_sz}
\end{figure}

While the BPU transiently overlooks the presence of \Bev branches, once these evicted branches retire, the BPU corrects the BHB value and flushes the pipeline to recover the correct state.
Therefore, unlike Spectre-BSE which has a persistent effect on branch history, the speculative execution of \Bipred{} and any leakage operations must occur \emph{before} all data and control dependencies of \Bev{} are satisfied and before the branches are resolved.
We observe that barrier instructions (\texttt{dsb isb} or \texttt{mfence}) placed before \Bipred effectively prevent speculating \tleak under vulnerable configurations, further confirming the speculation window requirements.

This timing constraint is crucial to exploit the speculative path effectively.
As illustrated in Fig.~\ref{fig:spec_bhs_window_sz}, our experiments demonstrate that when induce the mis-speculation through BTB/PHT eviction, Coretx-A76 and A78AE can execute more than 100 instructions within the speculative window, successfully performing the data leakage operation at its end.
Furthermore, if the BPU can detect a branch before its resolution (e.g., during the decoding stage), the speculation window may terminate prematurely.
Notably, we also observed that mistraining techniques yield speculation windows of comparable size, suggesting that the evaluated BPU implementations only detect the presence of a branch upon its architectural resolution rather than during earlier pipeline stages.



\section{Exploitation 3: BHS \& Fallback Predictions} \label{sec:chimera}

While the BPU updates the BHB with speculated outcomes, other architecturally resolving branches within the speculation window also influence predictions, potentially deviating from learned execution paths.
This section explores how this effect in BHS schemes can be exploited to truncate history-based predictions, demonstrated using \emph{extended Berkeley Packet Filter} (eBPF)~\cite{schulistLinuxSocketFiltering2024}.

\subsection{A Special Case of Legit eBPF Programs} \label{sec:eBPF-special}


Let us examine another code snippet consisting of two main blocks: the latter performs a data-dependent load, while the former initializes the register context that can repurpose the latter block as a Spectre gadget.
To prevent data disclosure through architectural execution, the snippet employs two \texttt{if} statements (conditional branches), denoted as {\bf \Bcinit} and {\bf \Bcload}, that are mutually exclusive through complementary conditions (i.e., \texttt{Bc\_init="if(flag)"} and \texttt{Bc\_load="if(!flag)"}).
This complementary structure ensures these blocks never execute together architecturally in the same instance. 
Additional branches may exist in implementations of such snippets to handle supplementary logic.

In eBPF, the \emph{verifier} statically examines each potential execution path in submitted programs to identify any violations of safety constraints. Among its various strict safety checks, the verifier enforces memory safety through two key requirements: first, all memory accesses must refer to a base pointer of a pre-allocated buffer, with the actual pointer value fixed at JIT compilation time; second, any added offset must be a scalar value within the buffer's size limits in any branch path.
Since the verifier evaluates these mutually exclusive blocks as separate execution paths and confirms their individual safety properties, such programs are deemed memory-safe and approved for loading. 

For programs with the structure introduced in Sec.~\ref{sec:eBPF-special}, exploitation would be straightforward in an environment vulnerable to out-of-place Spectre-v1.
Since the attacker has full control over the execution contexts of mistraining branches, they can separately mistrain multiple branches while preparing appropriate BHB values, even when BTB/PHT is indexed using both Program Counter and branch history.

However, our experiments on ARM processors in Section~\ref{sec:btb-eviction} demonstrate that both out-of-place Spectre-v1 training and BHB eviction require a congruent BHB value to succeed.
This requirement poses significant challenges for setting up the attack, as determining the necessary BHB value through static analysis can be difficult in real world.
Moreover, as shown in Section~\ref{spec-bhs-e}, Straight-Line Speculation (SLS) can interfere with branch history generation. Specific attack configurations---such as triggering SLS on \Bcinit---may prevent the processor from speculatively executing \Bcload to reach the data disclosure gadget due to altered branch history.
Given these complexities, we limited our analysis to the general case without considering out-of-place Spectre-v1 and SLS effects.

\subsection{Breaking The Speculative Path}

The mechanism of BHS suggests that a typical Spectre-v1 attack can affect the speculation of all subsequent branches under the BHS scheme.
However, while this behavior creates a limited attack surface for controlling subsequent speculation, it also impedes the creation of speculative flows that combine code blocks from different execution contexts.
Hence, a critical question emerges: \emph{is it possible to construct a code snippet that induces branch misprediction by exploiting history-based path speculation itself?}

Let us re-examine the behavior of branch history updating.
When both branches introduced in Sec.~\ref{sec:eBPF-special} appear within the speculation window, the speculated outcome of \Bcinit immediately updates the BHB, thus influencing the prediction of \Bcload.
However, if an attacker can force \Bcload to be \emph{predicted without using global branch history}, then \Bcload's prediction may become independent of \Bcinit's outcome.
This could enable combining elements from different legitimate flows into a single, BHS-induced speculative execution.
While history-based prediction can improve accuracy in most situations, BPUs must maintain the ability to predict based solely on Program Counter value to achieve better coverage in complex environments, particularly for branches that correlate poorly with history.
Many state-of-the-art BPU designs, such as TAGE~\cite{seznecCasePartiallyTagged2006,seznec201164,seznecNewCaseTAGE2011,yavarzadehHalfHalfDemystifyingIntels2023,seznec64KbytesITTAGEIndirect2011}, have implemented both PC-based and history-based sub-predictors (see Sec.~\ref{sec:branch-pred}).

\paragraph{Updating TAGE upon mis-prediction.}

TAGE predictors always update based on the encountered outcome of branches.
First, they update the provider component, which is the table that supplied the final prediction.
Then, upon a misprediction, if the provider component is not the table with the longest branch history, the BPU may allocate new entries in tables with longer histories, recognizing that the branch might correlate better with a longer history pattern.

For previously \emph{non-executed} branches, we thus hypothesize that the fallback mechanism selects the base predictor \texttt{T0} as the provider component, since all other tables report query misses.
The branch's outcome is then recorded in \texttt{T0}, establishing a new prediction entry with zero-length history.

\subsection{Attack Flow \#3b: Summon the Chimera from Fallback Predictions}\label{sec:chimera-ebpf}
While BHS limits the construction of in-place Spectre attacks on the programs of Sec.~\ref{sec:eBPF-special}, fallback behaviors in BPUs create an opportunity: branch predictions may not always depend on the speculated path, enabling BHS for execution paths that never existed architecturally from fragments of legitimate ones.
We demonstrate this variant of Spectre-BHS attack with an eBPF program that complies with the layout of Sec.~\ref{sec:eBPF-special}.

\paragraph{Branch history shuffling.} Similarly to the observations made by Wikner and Razavi in~\cite{wiknerBreakingBarrierPostBarrier2024} for x86\_64 architectures, while creating a nested speculative execution environment, we note that it is possible to ``shuffle'' the branch history related to \Bcload, forcing the BPU to \emph{make predictions independent of branch history}. A simple way to do so is to provide a dedicated conditional branch, say \emph{BHB-shuffle}, that is never architecturally taken before the attack so that it never updated branch history. Conversely, BHB-shuffle is intentionally taken at the stage of attack, 
hence injecting a history that was never encountered before.
This causes prediction to fall back to the \texttt{T0} predictor, which is also required to be trained.

\begin{algorithm}[th!]
    \SetKwFunction{set_secret}{queryBST}
    \SetKwFunction{UpdateBST}{updateBST}
    \SetKwFunction{Exit}{exit}
    \SetKwFunction{NOP}{NOP}
    \SetKwFunction{Memload}{memload}
    \SetKwData{params}{params}
    \SetKwData{takesc}{take\_sc}
    \SetKwData{esc}{esc}
    \SetKwData{setptr}{set\_ptr}
    \SetKwData{shufflebh}{shuffle\_BH}
    \params $\gets$ \emph{LEGIT\_PARAMS}\;
    \If(\label{ebpf_shortcut}){\takesc is FALSE}
    {
        \If(\tcp*[f]{\texttt{Bc\_init}}\label{ebpf_spec_win}){\setptr is TRUE}
        {
            \params $\gets$ \emph{\&SECRET}\;\label{ebpf_set_params}
        }
        \If(\label{ebpf_pred1}){\setptr is TRUE \& \esc is TRUE}
        {
            \Exit\;
        }
        \lIf(\label{ebpf_shuffle_bh}){\shufflebh is FALSE}{\NOP}
    }
    \If(\label{ebpf_pred2}){\esc is FALSE}
    {
      \Exit\;
    }
    \If(\tcp*[f]{\texttt{Bc\_load}}\label{ebpf_load}){\setptr is FALSE}
    {
        \Memload(\params)\;
    }
  \caption{A vulnerable program passing the eBPF verifier.}
  \label{alg_ebpf_poc}
\end{algorithm}

\paragraph{Exploitable snippet.}
We demonstrate a vulnerable snippet in Alg.~\ref{alg_ebpf_poc} that satisfies the conditions for a successful attack.
Besides \Bcinit and \Bcload, given that the discussed layout and eBPF verifier allow additional branches while 
preserving our requirements, we introduce some additional components to make the snippet practically exploitable.

\begin{enumerate}
\item
A conditional \emph{BHB-shuffle} branch (line~\ref{ebpf_shuffle_bh}) between \Bcinit and \Bcload meant to force fall-back predictions with \texttt{T0} when taken. This splits speculative execution into two parts: \textbf{(i)} the \emph{history-based part}, where branches are predicted using BHB, and \textbf{(ii)} the \emph{PC-based part}, where we will induce the BPU to predict using \texttt{T0} only (i.e., using PC values).
\item
Conditional escape blocks redirecting control flow outside the snippet both before and after the BHB-shuffle branch. The first escape (line~\ref{ebpf_pred1}) provides an execution path where the BHB-shuffle branch never executes architecturally, while the second one (line~\ref{ebpf_pred2}) allows branches in the PC-based part to avoid training the BPU. 
\item
A shortcut path to the PC-based part. This branch (line~\ref{ebpf_shortcut}) bypasses the history-based part, enabling isolated training of \texttt{T0} for branches in the PC-based part.
\end{enumerate}

Note that due to diverse microarchitectural implementatons and behaviors in real-world processors, other snippet structures may also be vulnerable to similar in-place mistraining.

\paragraph{Preparation.}

We initialize exploited BTB/PHT records using the following two flows.
During these training flows, we ensure the BHB-shuffling branch remains not-taken by setting \textsf{shuffle\_BH}=FALSE:
\begin{enumerate}[label=(\bf\Alph*)]
\item \label{flow_early_exit}
{ \textsf{take\_sc}=FALSE, \textsf{esc}=FALSE, \textsf{set\_ptr}=TRUE.}
\item \label{flow_fast_path}
{ \textsf{take\_sc}=TRUE, \textsf{esc}=TRUE, \textsf{set\_ptr}=FALSE.}
\end{enumerate}
Their branch traces are available in Appendix~\ref{app:chimera-traces}.
Our attack aims to mis-speculate a \emph{crafted} execution flow combining the \emph{history-based part} of Flow \ref{flow_early_exit} with the \emph{PC-based part} of Flow \ref{flow_fast_path}.
The PC-based part of Flow \ref{flow_fast_path} skips the escape command (\textsf{esc}=TRUE) and transmits data using a side channel (line~\ref{ebpf_load}). Since this part is never architecturally observed in the same execution flow together with taken branches from the preceding lines, it will leverage the base predictor \texttt{T0} for branch speculation.
To prevent TAGE from escalating to longer history-based predictions, we invoke this flow before any other training and avoid executing these branches in different contexts, ensuring these prediction records are created and remains in \texttt{T0}.

The history-based part of Flow \ref{flow_early_exit} initializes registers to make pointers dereference a secret address.
Since the snippet always executes through a common entry point (where we assume consistent branch history), we must train the BPU to speculate Flow \ref{flow_early_exit} under the default history-based prediction scheme, ensuring execution of register initializations (line~\ref{ebpf_set_params}).



\paragraph{Triggering data disclosure.}
We construct a vulnerable context by setting \textsf{shuffle\_BH}=TRUE to start the attack.
The attacker flushes \textsf{set\_ptr} from cache, then invokes the snippet with a {\bf dedicated attack flow}, with \textsf{take\_sc}=FALSE, \textsf{set\_ptr}=TRUE, and \textsf{esc}=TRUE to trigger the data leakage.

Due to the cache miss on line \ref{ebpf_spec_win}, the processor opens a speculation window and executes subsequent branches based on learned predictions.
Since the BPU is trained to speculate Flow \ref{flow_early_exit} at the common entry, line \ref{ebpf_spec_win} will be predicted as \emph{not taken} and line \ref{ebpf_pred1} as \emph{taken}.
When the execution flow reaches line \ref{ebpf_shuffle_bh}, since \textsf{shuffle\_BH} remains in the cache and the branch resolves as \emph{taken}, it leaves a footprint in the BHB.
From this point, subsequent branches encounter a previously unseen ``shuffled'' history. Based on the fallback prediction mechanism discussed earlier, all subsequent branches on lines \ref{ebpf_pred2} and \ref{ebpf_load} will be predicted using the PC-based \texttt{T0} base predictor.

Line \ref{ebpf_pred2} resolves quickly as taken since \textsf{esc} remains in the cache, thus skipping the escape opcode.
Finally, line \ref{ebpf_load} will use the PC-based prediction left by Flow \ref{flow_fast_path}, which is \emph{not taken}.
This triggers a memory load with the illegal, secret-dependent \textsf{params} set in line \ref{ebpf_set_params}.


In the end, the processor discovers its mis-speculation and reverts all speculative changes.
All speculative results, including the BHB-shuffling branch on line \ref{ebpf_shuffle_bh} and correlations among speculated branches, are not recorded by the BPU.
The branch on line \ref{ebpf_shuffle_bh} remains never-taken architecturally, and predictions for lines \ref{ebpf_pred2} and \ref{ebpf_load} stay unchanged.
To maintain an exploitable environment, the attacker must preserve and refresh the BTB/PHT entries for Flow \ref{flow_early_exit} and \ref{flow_fast_path} in their respective sub-predictors.



We first implement Algorithm \ref{alg_ebpf_poc} as a C program.
This program demonstrate successful speculation of both pointer setting and load gadget operations, achieving 100\% and 99.85\% success rates on Cortex-A76 and A78AE, respectively.

\newcommand{\mitxctx}{{\footnotesize \rotatebox[origin=c]{90}{\faAdjust}}}
\newcommand{\mitxpriv}{{\footnotesize \faAdjust}}
\newcommand{\mitxall}{{\footnotesize \faCircle}}
\newcommand{\mitnotapp}{{\footnotesize \faMinus}}

\newcommand{\expok}{{\footnotesize \faCheck}}
\newcommand{\expmit}{{\footnotesize \faTimes}}
\newcommand{\expno}{{\footnotesize \faMinus}}

\begin{table*}[ht]
\centering
\begin{tabular}{l|ccc|ccccc}
\hline
\multicolumn{1}{c|}{\multirow{2}{*}{\textbf{\emph{\(\mu\)Arch}}}} & \multicolumn{3}{c|}{\textbf{\emph{Mitigation}}}     & \multicolumn{4}{c}{\textbf{\emph{Primitive}}} \\
\multicolumn{1}{c|}{}     & BHB Clear   & CSV2     & BPU Flush & BHB    & BSE/BiasScp. & BHS    & Chimera   \\ \hline
Cortex-A72                &\mitxpriv    &\mitxall  &\mitxctx   &\expmit & \expok  & \expok & \expno    \\
Cortex-A76/A78AE          &\mitxpriv    &\mitxall  &\mitxctx   &\expmit & \expno  & \expok & \expok    \\ \hline
\multicolumn{1}{c|}{}     & BHI\_DIS\_S & e/aIBRS  & IBPB      & BHB    & BSE/BiasScp. & BHS    & Chimera   \\ \hline
Zen4                      &\mitnotapp   &\mitxpriv &\mitxctx   &\expmit & \expno  & M+C \& E+C & \expok    \\
Gracemont                 &\mitxpriv    &\mitxpriv &\mitxctx   &\expmit & \expno  & M+C   & \expok    \\
Redwood Cove / Crestmont &\mitxpriv    &\mitxpriv &\mitxctx   &\expmit & \expno   & M+C   & \expok    \\ \hline
\end{tabular}
\caption{Spectre mitigation techniques and exploitability of proposed attack vectors. For \emph{mitigations}, \mitxpriv ~=Enabled for cross-privilege, \mitxctx ~=Enabled for cross-context, \mitxall ~=Enabled by default, and ``\mitnotapp'' Not applicable. For \emph{primitives}, \expok ~=exploitable with recommended mitigations, \expmit  ~=fully mitigated, and \expno ~=Not applicable or not exploitable on this architecture. 
For Spectre-BHS on x86 processors, ``\textbf{M+C}'' and ``\textbf{E+C}'' indicate using \textbf{M}istraining or \textbf{E}viction to hijack kernel \textbf{C}onditional branches, respectively.
}
\vspace{-8px}
\label{tab:mitigations}
\end{table*}
\paragraph{Evaluation in eBPF.}
Kirzner and Morrison~\cite{kirznerAnalysisSpeculativeType2021} demonstrated that although eBPF verification ensures memory safety in architectural execution paths, it cannot protect eBPF against speculative execution.
Their work showed how \emph{cross-address-space, out-of-place} Spectre-v1 can compromise this safety assumption.
Their work proposed enhanced verification for unprivileged user programs that rigorously examines memory access constraints across all potential speculative execution paths, even those that are unreachable.
We believe this patch, implemented in v5.13rc7 and subsequently backported by various distributions, prevented the execution of Chimera from unprivileged contexts.
Consequently, our testing was conducted in privileged mode to bypass these restrictions.

In the eBPF implementation, \texttt{LEGIT\_PARAMS} satisfies the verifier by initializing two registers with a legitimate buffer pointer and offset for the data load block, while \Bcinit sets these registers to zero and \texttt{\&SECRET}, respectively. 

We tested this program in privileged mode.
To confirm whether the BHS is enabled in kernel space, we intentionally mistrained the bias of \Bcinit and flushed all four flag variables.
We observed that \texttt{LEGIT\_PARAMS} is successfully encoded in the data cache when \Bcinit is biased toward taken, confirming the presence of the BHS scheme for privileged conditional branches.
Given that Spectre-v1 attacks are widely recognized as not fully mitigated and rely on ad-hoc software mitigations, we conjecture that conditional branches, including privileged ones, can be exploited in Spectre-BHS attacks as victims through the manipulation of other branches within their speculative execution paths.

Through the malicious configuration discussed above, it successfully leaks arbitrary kernel memory contents.
While mis-speculations occasionally fail, we find that restarting it causes the kernel to assign a new address to the JITed snippet, avoiding interference and stabilizing the attack.
Under optimal conditions, we achieve a leakage rate of 24,628 Bit/s on A76 using single-pass bit extraction with 100\% accuracy.

Moreover, we successfully replicated these results across evaluated AMD and Intel processors, though failed to reproduce this attack vector on A72, suggesting the absence of the fallback mechanism on this microarchitecture.

However, it is important to note that, while this experiment was conducted with some mitigations disabled, similar exploitable patterns likely persist in production environments lacking comprehensive ad-hoc protections.


\section{Mitigations}
In Table~\ref{tab:mitigations}, we provide a comprehensive overview of existing hardware and software mitigations against our branch prediction attacks on the tested processors.

ARM has introduced a software-based BHB populating sequence~\cite{armSpectreBHBSpeculativeTarget2023} to clear branch footprints generated in user mode.
However, our experiments have demonstrated that this approach fails to isolate the BHB updating policy between user mode and privileged mode, allowing our attack primitives to bypass this countermeasure in all tested ARM processors.

AMD's AutoIBRS on Zen4 disables execution of predicted targets for kernel indirect branches, while Intel's \texttt{BHI\_DIS\_S} disables history-based prediction in kernel space entirely~\cite{wiebingTrainingSoloLimitations2025}.
These mechanisms significantly constrain the exploitability of kernel-space indirect branch targets on x86 processors; however, our testing reveals that prediction of conditional branches remains unrestricted despite these mitigations, leaving potential conditional branch-based attack vectors exploitable.
Thus far, our efforts have not yielded positive results in this area.

Recent microarchitectures implement implicit predictor mode separation by incorporating additional context information into branch prediction records.
Notable implementations include ARM's CSV2 and Intel's eIBRS~\cite{intelSpeculativeExecutionSide}.
Our evaluation demonstrates that these features remain insufficient to prevent BTB and PHR manipulation through mistraining and eviction, which subsequently influence the BHB updating process.

Some processors adopt aggressive Spectre-v2 mitigations like IBPB (x86)~\cite{intelIndirectBranchPredictor2018,LinuxKernelDocSpectre} and BPU flush (ARM)~\cite{armTrustedFirmwareA21202018} that clear prediction records.
ARM applies them to pre-CSV2 processors (e.g. A72), while they are widely deployed across x86.
Although these mitigations could effectively neutralize our attacks by invalidating malicious BPU configurations, a full deployment may degrade system performance by more than 50\%~\cite{doebelRePATCH02025}. 
Hence, the deployment is typically restricted to user space context switches, leaving syscalls unprotected.


\section{Conclusion}
This paper investigated how resource sharing and contention in modern BPUs
can originate security vulnerabilities in speculative execution when injecting 
inaccurate branch history. Our findings allowed to propose three novel attacks, 
Spectre-BSE, Spectre-BHS, and BiasScope, which were successfully tested on multiple processors, exhibiting a very high signal-to-noise ratio. 
A variant of Spectre-BHS was implemented by means of eBPF, demonstrating its capability of leaking kernel memory contents at 24,628 bit/s.
In the light of recent work~\cite{wiebingInSpectreGadgetInspecting2024,johannesmeyer2022kasper,qi2021spectaint,oleksenkoHideSeekSpectres2023} that revealed a wide availability of Spectre gadgets in the Linux kernel and 
the threats posed by speculative trojans~\cite{zhangExploringBranchPredictors2020},
this research should set the stage for future investigations to prevent these new attacks in uncontrolled environments.

\newpage

\section*{Acknowledgments}
We would like to thank the anonymous reviewers for their valuable feedback. This work was partially supported by project SERICS (PE00000014) under the MUR National Recovery and Resilience Plan funded by the European Union - NextGenerationEU, and by the Dottorato di Ricerca Nazionale in Cybersicurezza.

\section*{Ethics Considerations}
We disclosed our findings to ARM in September and October 2024. Among all the disclosed issues, ARM issued CVE-2024-10929 for Spectre-BSE, released a security advisory to address this vulnerability, and confirmed that BiasScope depends on the same underlying behaviour with it. We also disclosed our findings to Intel and AMD in November 2024.

We followed best practices for responsible disclosure, notifying ARM of
our findings as soon as possible, and keeping our findings confidential.
No experiments were performed with live systems. 

\section*{Open Science}
The artifacts implementing proof-of-concept demonstrations of Spectre-BSE, Spectre-BHS, BiasScope, and the Chimera attack are publicly available at \url{https://zenodo.org/records/15612187}.
These artifacts include comprehensive test modules, cross-context demonstrations, and the complete eBPF-based Chimera implementation.
We are confident that these artifacts enable the security community to verify and reproduce our findings.


\bibliographystyle{plain}
\bibliography{lib}

\appendix
\section{Cortex-A72: Determining Bias Status and Updating BST Records} \label{app:bias-algo}

Using the recorded data and the committed branch result, the BPU in Cortex-A72 determines the bias status of a branch via Alg.~\ref{alg_bias_determination} and excludes footprints from biased indirect branches during PHR updates.

\begin{algorithm}
  \SetKwFunction{QueryBST}{queryBST}
  \SetKwFunction{UpdateBST}{updateBST}
  \KwData{$outcome$ and $addr$ of the comitted branch}
  \KwResult{$biased$ status of the commited branch}
  $rec$ $\gets$ \QueryBST{$addr$}\;
  \eIf{$rec$ is NOT\_FOUND}
  {
    $biased$ $\gets$ TRUE\;
  }
  {
    \eIf{$rec.biased$ is TRUE}
    {      
      $biased$ $\gets$ ($rec.outcome$ = $outcome$)\;
    }
    {
      $biased$ $\gets$ FALSE\;
    }
  }
  \UpdateBST{$addr$, $biased$, $outcome$}\;
\caption{Determining the bias status of a branch.}
\label{alg_bias_determination}
\end{algorithm}

\section{Cortex-A76/A78AE: BTB/PHT Eviction and PC-Based Indexing} \label{sec:btb-ev-diff-hist}
Since BTB and PHT may employ multiple indexing mechanisms, we investigate the extent of branch history's contribution to index generation on Cortex-A76 and A78AE.
By modifying the branch history population parameters in mistrain snippets, we create scenarios where the BPU updates occur under branch history contexts that are different from the victim snippet.
Interestingly, while these modified mistrain snippets no longer achieve out-of-place mistraining, they still trigger BTB eviction when reaching the previously identified threshold on Cortex-A78AE.
This observation suggests that the BPUs in tested processors may employ both PC-based and history-based indexing schemes in different tables simultaneously.



\section{BTB/PHT eviction in canonical BHB} \label{sec:bhb-zen4}
\begin{figure}[t]
    \centering
    \begin{subfigure}{.5\linewidth}
      \includegraphics[width=\linewidth]{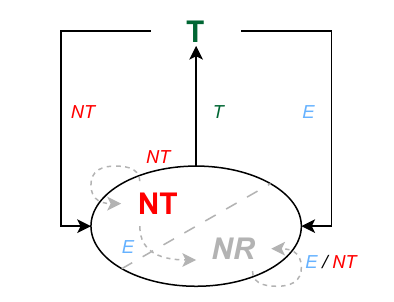}
      \caption{PHR}
      \label{fig_stat_trans_phr}
    \end{subfigure}%
    \begin{subfigure}{.5\linewidth}
      \includegraphics[width=\linewidth]{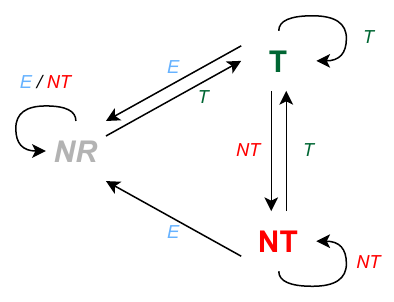}
      \caption{Canonical BHB}
      \label{fig_stat_trans_bhb}
    \end{subfigure}
    \caption{
    Transitions in BPU record status for a branch, induced by relevant branch instances.
    From the BPU's perspective, the status of a branch can be classified into three types: (i)~\textbf{``\textcolor[HTML]{006633}{T}''} for \emph{taken}, (ii)~\textbf{``\textcolor{red}{NT}''} for \emph{not taken}, and (iii)~\textbf{``\textcolor[HTML]{b3b3b3}{NR}''} for \emph{not recorded}. Transitions occur based on the recorded outcome (\textcolor{red}{NT} or \textcolor[HTML]{006633}{T}) or due to eviction (\textcolor{cyan}{E}).
    }
\label{fig_stat_trans}
\end{figure}
\paragraph{Never-taken branches in BHB.}
PHR implementations overlook undetected branches, effectively conflating ``not recorded'' and ``not taken'' states during speculative execution, while canonical BHB maintains a distinct ``Not Recorded'' default status for all branches and clearly differentiates between these states.
Our analysis on A72 reveals that conditional branches in the canonical BHB remain unrecorded until their first taken execution.
This observation aligns with AMD's explicit documentation that ``global history is not updated for not-recorded branches''~\cite{amdSoftwareOptimizationGuideZen42023}.

Based on these mechanisms, we can clearly identify three distinct states for conditional branches in the BHB updating process, as illustrated in Fig. \ref{fig_stat_trans_bhb}.
In such a canonical BHB implementation, evicting a previously-taken branch's entry does not cause it to appear not-taken in subsequent speculations; rather, it resets the branch to its initial unrecorded state.

In the attack vectors previously discussed, when attempting to evict the record of the conditional branch \Biprobe after eviction, the BPU may base its prediction of \Bipred on an entry associated with a third control flow path, distinctly different from both \flowa and \flowb.
This third path requires specific preparation strategies for successful exploitation.
Moreover, this ``not-recorded'' state may persist until the branch is taken for the first time, creating a long-lasting effect on BHB updating mechanism that extends beyond a single speculation window.

\paragraph{Revealing \emph{not-recorded} state.}
To demonstrate the impact of this explicit ``not-recorded'' state in BHB, we constructed an experiment that preserves the core setup from Section~\ref{spec-bhs-e}.
In this experimental setup, we insert a speculation barrier between \Biprobe and \Bipred to ensure all preceding branches are resolved and properly update the BHB before \Bipred is speculated.
\tsafe is modified to emit a distinctive side-channel signalenabling clear differentiation from mis-speculation to \tleak.
\Biprobe is executed as taken at least once prior to any training or testing sequences, ensuring its proper registration by the BPU.
Following all training and eviction operations, we invoke a dedicated test flow in which \Biprobe is not taken and \Bipred jumps to a third architectural target \texttt{t\_arch}.

To assess potential mis-speculation of \Bipred, we employ high-precision CPU timers (e.g., \texttt{rdtsc} on x86 processors and \texttt{mrs reg, pmccntr\_el0} on ARM processors) to measure the branch latency of \Bipred.
When \Bipred is correctly predicted to \texttt{t\_arch}, we observe relatively low branch latency since no speculation rollback is required.

\paragraph{Evaluation.}
We evaluated this setup on Zen4.
When eviction succeeded, \Bipred consistently demonstrated correct prediction with minimal latency, with no detectable side-channel signals from either \tleak or \tsafe.
This indicates that following \Biprobe eviction from BTB/PHR, the test flow generates a unique BHB value and associates it with \texttt{t\_arch}.
Even when \Biprobe architecturally resolves as not taken and all preceding branch outcomes match \flowa, the BPU predicts \Bipred using a third distinct state where \Biprobe is omitted due to its not-recorded classification.
However, this implicit ``bias'' status handling for conditional branches differs significantly from the bias-free scheme observed for indirect branches on A72.
Our analysis reveals that the canonical BHB implementation on A72 does not apply the bias-free scheme to conditional branches, which are always recorded in the 8-slot BHB even when consistently taken, following a distinctly different mechanism than that applied to indirect branches.

For conditional branch prediction, since only two possible predictions exist (taken or not taken), our experiments could not systematically demonstrate its perturbation effect.
However, inducing an unexpected BHB value may force the BPU to make PC-based predictions using fallback prediction, resembling the phenomenon discussed in Section~\ref{sec:chimera}.

\section{Execution Traces of Chimera}\label{app:chimera-traces}
The execution traces of training and attacks flows of the eBPF program 
used in Chimera (Section~\ref{sec:chimera-ebpf}) are reported in Table~\ref{tab_ebpf_bpu}.

\begin{table}[th!]
    \centering
    \begin{tabular}{lccccccccc}
    \hline
    \multicolumn{1}{l|}{\textit{Flow}}    & \multicolumn{3}{c|}{\textbf{A}}                 & \multicolumn{3}{c|}{\textbf{B}}                 & \multicolumn{3}{c}{\textbf{Attack}} \\
    \multicolumn{1}{l|}{\textit{Inputs}}  & F        & F      & \multicolumn{1}{c|}{T}      & T        & T      & \multicolumn{1}{c|}{F}      & F          & T          & T         \\ \hline
    \multicolumn{1}{l|}{\textbf{Line 2}}  & \multicolumn{3}{c|}{\cellcolor[HTML]{FFFFC7}NT} & \multicolumn{3}{c|}{TT}                         & \multicolumn{3}{c}{NT}              \\ \hline
    \multicolumn{1}{l|}{\textbf{Line 3}}  & \multicolumn{3}{c|}{\cellcolor[HTML]{FFFFC7}TT} & \multicolumn{3}{c|}{--}                         & \multicolumn{3}{c}{NT}              \\ \hline
    \multicolumn{1}{l|}{\textbf{Line 5}}  & \multicolumn{3}{c|}{\cellcolor[HTML]{FFFFC7}TT} & \multicolumn{3}{c|}{--}                         & \multicolumn{3}{c}{NT}              \\ \hline
    \multicolumn{10}{c}{\textit{\small    Split by BHB-shuffle branch on  Line 7}}                                                                                                          \\ \hline
    \multicolumn{1}{l|}{\textbf{Line 8}}  & \multicolumn{3}{c|}{NT}                         & \multicolumn{3}{c|}{\cellcolor[HTML]{FFFFC7}TT} & \multicolumn{3}{c}{--}              \\ \hline
    \multicolumn{1}{l|}{\textbf{Line 10}} & \multicolumn{3}{c|}{--}                         & \multicolumn{3}{c|}{\cellcolor[HTML]{FFFFC7}NT} & \multicolumn{3}{c}{--}              \\ \hline
\end{tabular}
\caption{
    Architectural execution traces of training and attack flows. Three traces shown: Flow \ref{flow_early_exit}, Flow \ref{flow_fast_path}, and the attack flow.
    Input variables (\textsf{take\_sc}, \textsf{esc}, \textsf{set\_ptr}) are shown in the header, with ``T'' indicating TRUE and ``F'' indicating FALSE. For each branch, identified by line number, ``TT'' denotes taken, ``NT'' denotes not-taken, and ``--'' denotes not executed.
    Branch outcomes used to craft the malicious speculative execution path are highlighted in yellow.
}
\label{tab_ebpf_bpu}
\end{table}
\end{document}